\documentstyle[11pt,IAUS215,twoside,epsfig]{article}

\def\edcomment#1{\iffalse\marginpar{\raggedright\sl#1\/}\else\relax\fi}
\reversemarginpar

\begin{document}
\vspace*{1cm}
\title{The Rotation of Low-Mass Pre-Main-Sequence Stars}
 \author{Robert D. Mathieu}
\affil{University of Wisconsin, Department of Astronomy, Madison WI 53706  USA}

\begin{abstract}

Major photometric monitoring campaigns of star-forming regions in the past decade have provided rich rotation period distributions of pre-main-sequence stars. The rotation periods span  more than an order of magnitude in period, with most falling between 1 and 10 days. Thus the broad rotation period distributions found in 100 Myr clusters are already established by an age of 1 Myr. The most rapidly rotating stars are within a factor of 2-3 of their critical velocities; if angular momentum is conserved as they evolve to the ZAMS, these stars may come to exceed their critical velocities. Extensive efforts have been made to find connections between stellar rotation and the presence of protostellar disks; at best only a weak correlation has been found in the largest samples. Magnetic disk-locking is a theoretically attractive mechanism for angular momentum evolution of young stars, but the links between theoretical predictions and observational evidence remain ambiguous. Detailed observational and theoretical studies of the magnetospheric environments will provide better insight into the processes of pre-main-sequence stellar angular momentum evolution.

\end{abstract}

\section{The Angular Momentum Problem}

The specific angular momenta of the interstellar clouds from which stars form 
are significantly higher that that of the Sun, a fact that has been recognized for 
many decades. The numbers shown in Table 1 give scale to the ``angular 
momentum problem", with dense molecular cores having six orders of 
magnitude higher specific angular momentum than the Sun. In many, and 
perhaps all, cases the clouds partially solve this problem by depositing much of their 
angular momentum in the orbital motions of stellar and planetary companions. 
As shown by Vogel \& Kuhi (1981) and Hartmann et al. (1986), most of the angular momentum 
problem has been solved by the pre-main-sequence  (PMS) stage of evolution. Nonetheless, a factor 10-100 reduction in specific 
angular momentum is still required to derive the solar rotation at 4.5 Gyr, and indeed to produce the 
rotation of main-sequence solar-mass stars at 100 Myr.

\begin{center}
\begin{tabular}{ll}
\multicolumn{2}{l}{{\bf Table 1. Specific angular momenta (cm$^2$s$^{-1}$)}} \\ 
Dense molecular cores & $10^{21-22}$ \\
Wide binary stars &  $10^{19-20}$ \\
Pre-main-sequence stars & $10^{16-17}$ \\
Sun &10$^{15}$ \\
\end{tabular}
\end{center}

\begin{figure}
\begin{center}
\epsfxsize=4in
\epsfbox{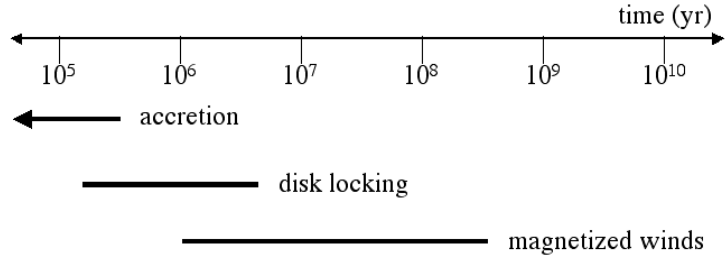}
\caption{Schematic timeline of external torques suggested to be acting on solar-mass single stars (courtesy of Prof. Terndrup).}
\end{center}
\end{figure}

Thus many recent investigations into the rotation of low-mass stars, both theoretical 
and observational, have focussed on identifying the torques modifying stellar angular momentum from protostars to the Sun. Figure 1 shows a schematic of those 
mechanisms presently thought to be in play and the timescales on which they 
operate. During the PMS phase much recent work has been done within the context of the magnetic disk-locking 
mechanism developed by Shu et al. (1994). Here the ionization fraction in a
protostellar disk is taken to be sufficient to couple the stellar magnetic field to the disk. At a certain 
radius in the disk, set by the stellar magnetic field strength and the mass accretion rate, the 
magnetic torque on the disk material balances the viscous torque. The 
equilibrium configuration is one where the star is in corotation with 
Keplerian rotation at that radius. Higher mass accretion rates drive that radius 
toward the star, leading to more rapid stellar rotation and shorter rotation periods. 
Higher magnetic field strengths move the radius away from the star, leading to 
slower stellar rotation and longer rotation periods. Importantly, in this picture 
the rotation period is independent of the stellar radius. Thus, if the magnetic field 
and mass accretion rate are constant, the star remains "locked" at  a rotation 
period independent of its PMS contraction.

In this paper we review recent observational studies of PMS stellar rotation, with particular emphasis on their implications for this disk-locking model of  early angular momentum evolution.

\section {Rotation Period Distributions of Pre-Main-Sequence Stars}

The basic data for stellar angular momentum studies are rotation period 
distributions. PMS stars have large star spots which, as the stars rotate, modulate the flux observed at Earth by as much as several percent. Thus periodicities derived from light curves provide a very 
accurate ($\approx$ 1\%) measure of stellar rotation periods. Indeed, stellar rotation period is the most accurate physical parameter that we know for PMS stars.

\begin{figure}[ht]
\begin{center}
\epsfxsize=3in
\epsfbox{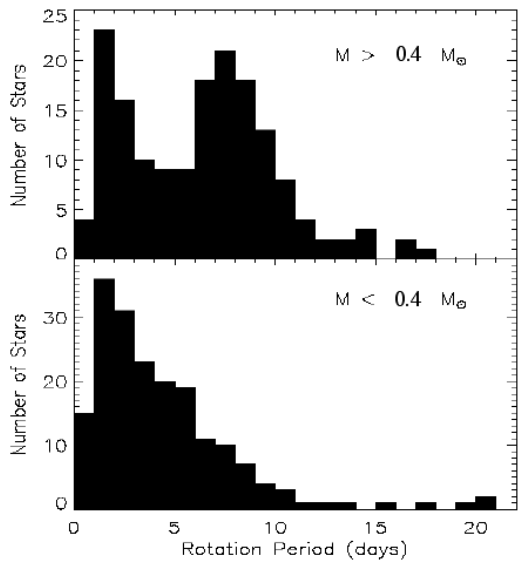} 
\caption{Stellar rotation period distributions for stars in the core of the Orion Nebula Cluster, divided by stellar mass at 0.4 M$_{\odot}$. (Adapted from Herbst et al. 2002, using mass calibration of Siess et al. 2000.)}
\end{center}
\end{figure}

A great deal of observational work has focussed on the Orion Nebula Cluster 
(ONC), since the high surface density of PMS stars allows CCD imaging 
to yield large numbers of rotation periods (e.g., Choi \& Herbst 1996, Stassun et al. 
1999, Rebull 2001, Herbst et al. 2002). The total count is over 1000, with 
 $\sim$ 40\% of all ONC stars yielding rotation periods. (See summary of Herbst in Stassun \& Terndrup 2003.) The most recent ONC rotation period distributions are shown in 
Figure 2, as derived by Herbst et al. (2002). Most of the rotation periods are 
found to be between 1 and 10 days. There has been some concern that the marked 
decrease in number of rotation periods longer than 10 days is a selection effect resulting 
from reduced dynamo activity and spot sizes in slow rotators, but vsin i studies tend 
to argue against this (Rhode, Herbst, \& Mathieu 2001).

Bimodality in the ONC period distribution has been the subject 
of extensive discussion in the recent literature. A bimodal structure is present at a statistically significant level
for stars with masses greater than 0.4 M$_{\sun}$ in the central region of the ONC 
(using the mass calibration of Siess, Dufour, \& Forestini 2000). However, in the regions 
flanking the cluster core, Rebull (2001) does not find a bimodal distribution 
among stars of similar numbers, mass and age (specifically, similar location in 
the HR diagram), a difference that remains to be resolved. The importance of 
resolution lies in the attribution of the long-period peak ($\sim$ 8 days) in the bimodal distribution to slowly 
rotating, disk-locked stars, while more rapidly rotating stars have been suggested to have 
spun up as a result of PMS contraction after the disk lock is lost (Bouvier et al. 1993, Choi 
\& Herbst 1996).

Perhaps as significant as the bimodality itself is that the range of rotation 
periods extends over an order of magnitude. With the discovery in the early 1980's of 
rapidly rotating low-mass stars in the Pleiades, a challenge to theorists has been 
to create a broad distribution in stellar rotation at 100 Myr (see the excellent review of past work by Barnes, Sofia \& Pinsonneault 2001 as well as in these proceedings). In fact, the {\it relative} width of the period distribution in the ONC at 1 Myr already mimics that in the Pleiades, as shown in Figure 3. Recently Tinker, Pinsonneault, \& Terndrup (2001) have modeled the evolution of the ONC distribution to 100 Myr, with some success in recreating the observed Pleiades rotation period distribution. There remains much work to be done, however, as
these models use a parameterized approach to the braking mechanisms, and rather long disk-locking lifetimes are required (3-8 x 10$^6$ yr).

Importantly, the shape of the PMS rotation period distribution depends upon stellar mass (Figure 2; Herbst et al. 2000, 2002). The period distribution for stars with masses less than 0.4 M$_{\sun}$ is substantially more concentrated toward shorter rotation periods than the period distribution for higher mass stars. Interestingly, this mass dependence is also seen in the Pleiades at 100 Myr (Terndrup et al. 2000); once again the basic structure of the older rotation period distribution is established by 1 Myr.

\begin{figure}[ht]
\begin{center}
\epsfxsize=3in
\epsfbox{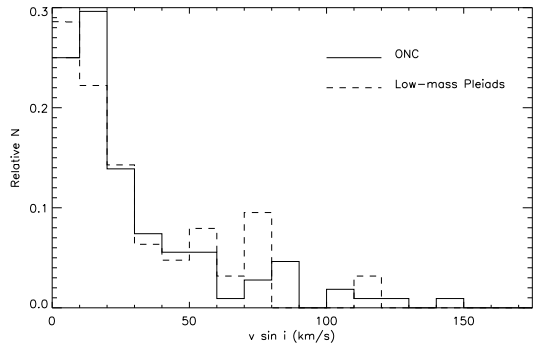}
\caption{Comparison of the vsin i distributions for low-mass stars in the ONC and the Pleiades, demonstrating that the relative width of the Pleiades rotation period distribution has already been established by the 1 Myr age of the ONC. (Note that the two stellar populations have different radii, so the similarity of the vsin i distributions does {\it not} imply similar rotation period distributions.) (Taken from Stassun et al. 1999.)}
\end{center}
\end{figure}

\section{Pre-Main-Sequence Stellar Rotation and Protostellar Disks}

Considered simply, the disk-locking model might be expected to predict a 
prevalence of protostellar disks associated with the most slowly rotating PMS 
stars (i.e., the stars that are disk-locked) and a lack of disks associated with more 
rapidly rotating PMS stars (i.e., the stars no longer locked which have spun up 
with their contraction). This is in fact what Edwards et al. (1993) found in their initial study of
Taurus and ONC PMS stars, using 2 $\micron$ excess emission as their diagnostic for the 
presence of disks.

\begin{figure}[ht]
\begin{center}
\epsfxsize=3in
\epsfbox{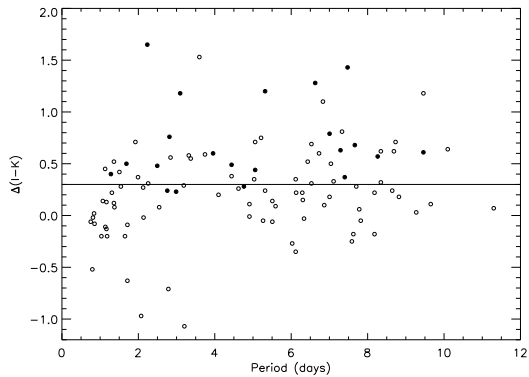}
\caption{Plot of 2 $\micron$ excess $\delta$(I-K) vs. stellar rotation period for ONC stars. A $\delta$(I-K) greater than 0.3 is indicative of 2 $\micron$ continuum emission from a circumstellar disk (Hillenbrand 1997). Those stars represented by filled circles also show emission-line evidence for an accretion disk. (Taken from Stassun et al. 1999).}
\end{center}
\end{figure}

Thus it was surprising when Stassun et al. (1999) found no correlation between 2 
$\micron$ excess and stellar rotation period among the ONC PMS stars (Figure 4). Stassun 
et al. also used H$\alpha$ and/or Ca triplet emission line strength (filled circles in Figure 4) as a 
diagnostic for active mass accretion, again finding no correlation with rotation 
period. The same conclusion was drawn by Rebull (2001) using ultraviolet 
excess as a diagnostic for mass accretion. More recently Herbst et al. (2002) have argued for a weak but significant
correlation of 2 $\micron$ excess with rotation period for ONC stars both above and below 0.4 M$_{\sun}$. Even so, in their sample many rapid rotators show accretion disk diagnostics, while many slow rotators do not.
A simple association of protostellar disks with slow PMS stellar rotation is not present in the ONC.

Importantly, the predictions of disk-locking theory for emission excesses are not yet secure. For example, disk-locking theory predicts 
that gaseous disks will not exist within the corotation radius, since 
accretion of disk material proceeds along the magnetic field lines to the stellar surface. For a stellar 
magnetic field strength of 1 kG, a mass accretion rate of 10$^{-7}$ M$_{\sun}$/yr, 
and a stellar mass of 0.5 M$_{\sun}$, the corotation radius lies at 14 R$_{\sun}$. Using simple disk emission models, Stassun et al. (2001) showed that a disk with such a large central hole may not
show a 2 $\micron$ excess. Indeed, in their models the peak 2 $\micron$ excess derives 
from a disk
with a central hole of radius 6 R$_{\sun}$, which corresponds to a rotation period of 2 
days for a 0.5 M$_{\sun}$ star. Thus, if disk locking is in force, the predicted correlation of 2 $\micron$ excess with rotation period would be in the opposite sense of that found by 
Edwards et al. and Herbst et al.

This simple model for disk structure and emission is meant only to demonstrate 
that the observational predictions of disk-locking theory for excess emission are not yet clearly 
established. Furthermore, if we are to continue to use 2 $\micron$ 
excess as our disk diagnostic, we must develop a better understanding of the dynamics and emission 
properties of the dust. For example, Shu (private communication) has pointed out that while 
the gas will follow the magnetic field lines, dust may diffuse into the central hole. 
Alternatively, the emission from dust traveling along stellar field lines is not yet known; 
presumably this will depend greatly on the surface area of the flux tubes. 

Significantly, Stassun et al. (2001) used 10 $\micron$ excess to search for disks 
among PMS stars of all rotation periods showing no 2 $\micron$ excess. In only three cases among 32 stars was 10 $\micron$ disk emission found in the absence of near-infrared emission. Evidently, slow and 
rapid rotators without 2 $\micron$ excess in fact do not have protostellar disks 
to brake the stars. 

More fundamentally, the disk-locking model does not in itself predict a preferred rotation period; the rotation period of a given star is set by the combination of stellar magnetic field, mass accretion rate, and stellar mass. Thus it is not evident that all disk-locked PMS stars should have similar rotation periods. Neither is it evident that all disk-locked PMS stars should rotate slowly. Indeed, given the certainty that at least the mass accretion rates decrease steadily from the protostar phase to the late-PMS phase, one might expect that, given disk locking, stellar rotation would evolve from rapid to slow (due to decreasing mass accretion rate) to rapid again (after the release of disk locking and subsequent stellar contraction). The findings of Greene \& Lada (2002) that protostars are more rapidly rotating than PMS stars are compatible with this conjecture. Of course, the stellar magnetic fields themselves also may vary significantly over the same timescales, and both mass accretion rates and stellar magnetic fields will vary from star to star at any given time. So a strong correlation of stellar rotation with the simple presence of disks would not be expected. Perhaps the weak correlation found by Herbst et al. (2002) might be explained by the one-directional trend of any star to spin up once disk-locking ends.

Finally, it is worth noting that mass accretion rates have been observed to vary stochastically on short timescales, and if FU Ori events are part of every PMS star's history the accretion rates also have large excursions on long timescales. Indeed such events of high mass accretion rates might act to spin up PMS stars. Critically, the response time of stellar rotation to variations in the disk-locking radius must be explored. Hartmann (2002) suggests that the response times are quite long ($\sim$ 10$^{6-7}$ yr), perhaps so much so that the effectiveness of any disk-locking during the PMS phase is limited.

\section{Rapid Stellar Rotation}

\begin{figure}[ht]
\begin{center}
\epsfxsize=3in
\epsfbox{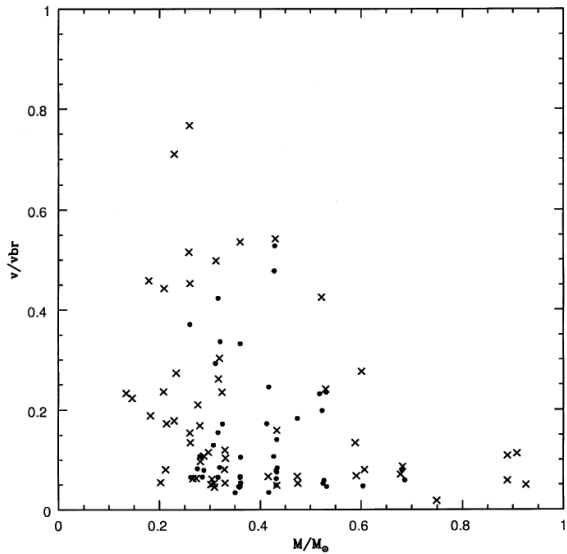}
\caption{Plot of stellar equatorial velocity normalized to critical velocity vs. stellar mass for ONC stars. The 'x' symbols represent stars without disks, based on 2 $\micron$ excess. (Taken from from Clarke \& Bouvier 2000).}
\end{center}
\end{figure}

With a high frequency of observation (1 hr$^{-1}$),  Stassun et al. (1999) were able 
to identify a number of very rapid rotators with sub-day periods. The shortest 
PMS rotation period discovered to date is 0.27 days. As shown in Figure 5, some of 
these rapid rotators are spinning within a factor of two of their critical velocities. Furthermore, ONC stars with masses of 0.2 - 0.5 M$_{\sun}$ will contract by
factors of several before reaching the ZAMS. Finally many of the most rapidly rotating stars do not 
show any evidence of associated protostellar disks.

Thus these stars pose an interesting challenge. In the absence of angular momentum loss these stars would exceed their critical velocity as they approach the ZAMS. In addition, they have no evident means for 
rapid braking over that contraction time, although winds will have some effect. 
Rebull (2001) has pointed out that the calculation of a star's critical velocity is dependent upon uncertain mass and radius determinations, and indeed the 
atmospheres and colors of these rapidly rotating stars may differ from standard 
models. With that caveat acknowledged, these stars 
give motivation for theorists to investigate the pre-main-sequence evolution of 
very rapidly rotating stars.

It may be physically significant that these rapidly rotating stars were discovered 
in the ONC and not in the earlier Taurus-Auriga studies,
where no sub-day rotation periods have been found. Clarke \& Bouvier 
(2000) compared the Taurus-Auriga and ONC period distributions and 
determined that they differ at a confidence level in excess of 99\%, in the sense that the ONC 
has a larger population of rapid rotators. Their conjectures for the origin of this 
physical difference include a larger population of tidally locked binaries in 
the ONC, weaker stellar magnetic fields in the ONC, a higher 
frequency of stellar encounters in the ONC cluster destroying disks and ending 
disk-locking at an earlier age, or differing initial conditions. As they discuss, none of these explanations are without difficulties. (As an aside, Stassun and Mathieu have underway a spectroscopic monitoring study of the ONC rapid rotators. As yet a high frequency of spectroscopic binaries have not been found.)

If this environmental difference stands the test of further observation, it may 
prove seminal for our understanding of the origins of stellar angular momentum. 
Most critical will be extensive rotation period studies in other Taurus-Auriga-
like regions so that the population differences can be put on more secure ground.

\section{ Pre-Main-Sequence Evolution of Stellar Angular Momentum}

As many modeling investigations have found, stellar winds alone do not seem capable of providing the angular momentum loss required by young stars before reaching the main sequence (e.g., Tinker et al. 2002). Thus most recent scenarios for PMS angular momentum evolution have the general form of strong braking due to disk-locking during the protostar phase and early in PMS evolution, followed by weak braking by winds after the loss of these disks. The external torques provided by winds are thought to be sufficiently weak as to permit substantial spin-up of the stars as they contract to the main sequence.

Such a scenario is, in principle, straightforwardly testable by observing the evolution of stellar rotation period distributions with age. This approach has been somewhat stymied by the paucity of rich clusters with ages between the ONC and the Pleiades (roughly 1 Myr to 100 Myr), but nonetheless observational work has begun. Perhaps befitting the early stage of the work, the first studies have not obtained findings consistent with either these scenarios or with each other. 

Rebull et al. (2002 and these proceedings) have analyzed both the rotation period and vsin i distributions in the core of the ONC, in the fields flanking the ONC, and in NGC 2264 (an older cluster with an age of roughly 2-4 Myr). They take stellar radius as a diagnostic of age for stars of a given spectral type. Here stellar radii are derived from luminosities and effective 
temperatures , while spectral type is taken as a diagnostic for stellar mass for stars on the Hayashi portion of their PMS evolution (i.e., evolution at nearly constant effective temperature). They find no evidence up to an age of 3 Myr for stars spinning up as they contract down the Hayashi track.

This observational result is essentially that expected from strong disk locking, where the rotation period is held fixed by coupling to the disk. However, this interpretation is problematic in that most of the stars in the sample lack 2 $\micron$ evidence for disks. Conceivably 2 $\micron$ excess is not a sufficiently sensitive diagnostic for disks (e.g., Lada et al. 2000), but as noted above Stassun et al. (2001) found very few cases of stars not showing 2 $\micron$ excess emission yet having disk emission at 10 $\micron$. 

In contrast to this result, Lamm et al. (these proceedings) find that the rotation period distribution in NGC 2264 is significantly different from that in the ONC. Specifically, they find that the peaks in both the higher- and lower-mass rotation period distributions are at periods roughly a factor of 2 shorter than found in the ONC. (Although, importantly, the relative shapes of the higher- and lower-mass distributions are very similar to the ONC, with the lower mass stars NGC 2264 stars again rotating more rapidly.) If between 1 Myr and 2-4 Myr angular momentum is conserved, then the more rapid rotation in NGC 2264 could be attributed simply to the contraction of the stars to smaller radii. Of course, in this case the conclusion would be that disk locking is not significant after 1 Myr, precisely the opposite of the indication from the Rebull et al. result.

Clearly we are at an early stage of this line of investigation, 
particularly since both Lamm et al. and  
Rebull et al. base their analyses on (independently derived)
NGC 2264 rotation periods. 
Equally clearly such observations have the potential to define stellar angular momentum evolution from the protostar to the ZAMS, and so must be pursued with vigor.

\section {Future Directions}

While the rich statistical studies have provided important foundations and insights, we now need direct attacks on the physics of angular momentum evolution during the protostar and PMS phases of stellar evolution. Within the context of testing the disk-locking model, future observations should include:

1) Direct observation of the kinematics and spatial distribution of {\it gas} in 
disks. New high-spectral-resolution infrared spectrographs are permitting 
observation of CO lines in disks. Not only are these gas tracers more 
sensitive to the presence of disks than dust emission, they provide direct spatial information 
on the emission sources under the presumption of Keplerian rotation. A 
recent study of CO fundamental lines by Najita, Carr, \& Mathieu (2003) 
suggested that the inner edges of circumstellar gas disks lie within the 
corotation radii, but these are only the first observations of a rich 
direction of study.

2) Extensive measures of mass accretion rates and stellar magnetic fields. Mass accretion rates have been 
under intense study for over a decade. Recently Johns-Krull \& Gafford (2002) 
have studied relationships predicted by disk-locking among stellar mass, radius, accretion rate and rotation period. Under the assumption that the magnetic field strength does not vary appreciably from star to star, they find support for the Shu et al. (1994) formulation of disk-locking, extended to include non-dipole field topologies. Magnetic field strengths of PMS stars have only recently been 
measurable (Johns-Krull et al. 2001), and will ultimately allow detailed tests
of whether inner disk structures are being molded by magnetic fields.

On the theoretical side, we need:

1) Dynamical studies of stellar rotation evolution in the presence of stellar
magnetic fields and accretion disks. Of particular interest will be timescales for rotation evolution given evolving and variable accretion rates. If magnetic coupling to accretion disks occurs, do young stars achieve equilibrium configurations? More generally, what are the evolutionary paths for stellar rotation from the protostar to the birthline to the main sequence? As briefly discussed in Section 3, the paths may be complex, with even the sign of the evolution changing with time. These studies will be a major 
computational challenge, but they are necessary in order to predict observables, such as stellar rotation distributions.

2) Fully consistent magnetohydrodynamic modeling across the boundary of the stellar surface. In recent studies either the braking mechanisms are parameterized 
and the stellar interior is carefully done (e.g, Barnes et al. 2001, Tinker et al. 2002) or the 
physics of braking is carefully done with the interior structures approximated 
(Collier Cameron \& Campbell 1993). Recently a new paradigm for low-mass stellar rotation 
developed by Barnes (2003) rather naturally explains the evolution of stellar  rotation distributions in main-sequence stars, including bimodalities 
and mass dependences observed in the period distributions. This model provides a 
framework for connecting internal and external magnetohydrodynamic processes. 
Ultimately these new ideas must interface PMS period distributions at 1 Myr, and the same magnetohydrodynamic issues addressed for PMS stars and younger.

\medskip
Every indication is that a strong braking mechanism is needed if the rotational properties of protostars and young PMS stars are to evolve into the rotational properties of main-sequence stars. Magnetic disk-locking is a theoretically attractive mechanism for such strong braking. Observational indications of active disk locking at ages of 1 Myr and earlier are beginning to appear, most notably the detailed tests of Johns-Krull \& Gafford (2002), the rapid rotation of protostars compared to PMS stars at 1 Myr, and perhaps the weak correlation of near-infrared excess with longer rotation period in the ONC. 
A critical outstanding question is the duration of disk locking, and in particular whether it can remain effective for several Myr into the pre-main-sequence phase of evolution.

\medskip
Acknowledgements - It is a pleasure to acknowledge the important contributions of many colleagues past and present to the results presented here, with particular thanks for long-standing collaborations and associations with Bill Herbst, Luisa Rebull, and Keivan Stassun.

\end{document}